# Nonparametric Clustering Stopping Rule Based on Multivariate Median

Hend Gabr[1,2], Brian H Willis[3], and Mohammed Baragilly[4,5]

[1] Department of mathematics, insurance and statistics, Faculty of Business, Menoufia University, Egypt.

[2] Alliance Manchester Business School, University of Manchester, Manchester, UK.

[3] Institute of Applied Health Research, University of Birmingham, UK.

[4] Department of Mathematics, Insurance and Applied Statistics, Helwan University, Egypt.

[5] Department of Inflammation and ageing, University of Birmingham, UK.

Correspondence should be addressed to Mohammed Baragilly; m.h.h.baragilly@bham.ac.uk

**Abstract**

This paper introduces a novel nonparametric criterion for determining the appropriate number of clusters. The method is constructed to reconcile two competing objectives of cluster analysis: the preservation of internal homogeneity within clusters and the maximization of heterogeneity across clusters. To this end, the proposed algorithm optimizes the ratio of inter-cluster to intra-cluster variability, incorporating adjustments for both the sample size and the number of clusters. Unlike conventional techniques, the method is distribution-free and demonstrates robustness in the presence of outliers. Its properties were first examined through extensive simulation studies, followed by empirical evaluations on three applied datasets. To further assess comparative performance, the proposed procedure was benchmarked against 13 established algorithms for cluster number determination. In 11 of these comparisons, the proposed criterion exhibited superior performance, thereby underscoring its utility as a reliable and rigorous alternative for multivariate clustering applications.

## 1. Introduction

The problem of determining the number of clusters in a dataset is a relevant problem across a wide variety of disciplines such as business [1], psychology [2], statistics [3], medicine [4], computer science [5], and engineering [6]. The clustering stopping rule (also called an index) focuses on computing some functions for each cluster solution and chooses the one that indicates the most distinct clustering. Many publications have introduced various parametric statistical algorithms for identifying groups of points in multidimensional spaces while considering within and between cluster variations depending on the classical parametric measures "mean" [7], [8]. The nonparametric measures like spatial medial and spatial rank represent another approach to determine the optimal number of clusters in multivariate datasets [9], [10], [11], [12].

In many statistical studies, a particular attention is paid to the spatial median due to its straightforward computational approach and its robustness against outliers where, unlike the mean, it has the highest possible breakdown point of 50%. Moreover, using the ranks instead of the original observations provides more information about how central each observation is and in which direction it is moving from the center. The term "spatial median" is presented by [13], who examined its properties and showed that the empirical distribution of the spatial median is asymptotically normal. Another property of the spatial median is its uniqueness whenever the dimension is two or more ($p \geq 2$) and can be extended it into Banach spaces [14]. Other properties of spatial median have been introduced by [15] who showed that although the spatial median is equivariant under orthogonal transformations, it is not equivariant under general non-singular transformation. Thus, one can use the corresponding transformation-retransformation spatial median that has been extensively discussed in [16] [17]. For both the spherical and general elliptic model, the transformation retransformation spatial median is efficient and performs well. Another affine equivariant spatial median has been introduced by [18] who introduced this equivariant version of spatial median by estimating the location and shape parameters simultaneously based on the spatial signs. Another fast and monotonically convergent algorithm in order to compute the spatial median is proposed by [19].

In this study we introduce a clustering stopping rule algorithm based on the spatial median that considers the within and between distances in the multidimensional space. We propose an algorithm aiming at achieving the trade-off between the homogeneity within the clusters and the heterogeneity between clusters while controlling for number of clusters as well as number of observations within each cluster. The homogeneity within the clusters implies that data belongs to the same cluster should be as similar as possible such that the closer the observations are to each other within a cluster, the more cohesive and well-defined the cluster is. Conversely, the heterogeneity between clusters implies that data belongs to different clusters should be as different as possible such that the greatest the distance between clusters, the more distinct clusters are. The rationale behind our method is that as the number of clusters increases, the within-cluster variation will decrease, and the between-cluster variation will initially increase. Thus, maximizing the ratio of the variation between clusters and the variation within clusters while adjusting for the number of clusters and number of observations achieves the balance between cluster separation and cohesion within clusters.

This method aligns with ideas behind common criteria like the Elbow method in clustering analysis, where the goal is to find a balance between too few and too many clusters [20].

We evaluated the stability and the efficacy of the proposed methods using both simulated and real-world datasets. Moreover, we compared the performance of our model with other traditional methods for determining the number of clusters.

## 2. Definitions

*2.1 The sample univariate median:*

Suppose that $x_1, \ldots, x_n$ is a sample from a univariate distribution $F$ and let $x_{(1)} \leq x_{(2)} \leq \cdots \leq x_{(n)}$ be the order statistics of this sample, then the univariate median is defined as the $\frac{1}{2}$th quantile of the underlying distribution. Alternatively, we can use the univariate sign function that has been defined in [20] in order to get the univariate median value, where the numbers of observations on the left and right of x is $|\sum_{i=1}^{n} sign(x_i - x)|$ which means that the univariate median is

$$\text{any } x \in \mathbb{R} \text{ that satisfies} \left|\sum_{i=1}^{n} sign(x - x_i)\right| = 0 \qquad (1)$$

Unlike the univariate median, the multivariate median has various versions and computational methods. For instance, $L_1$ –median, $L_2$ –median (spatial median), $L_p$ –median, half-space median, simplicial depth median, marginal median and Oja median are well known types of the multivariate median (see [21] for a comprehensive review of multivariate medians).

*2.2 The Spatial median:*

Suppose that $x_1, \ldots, x_n \in \Re^d$ where $d$ is the number of variables and n is the number of observations, considering the multivariate sign function that defined in [20] and replacing the absolute value $|\cdot|$ in univariate median above (1) with the Euclidean norm $\|\cdot\|$ then the spatial sample median can be written as

$$\text{any } x \in \Re^d \text{ that satisfies } \left\| \sum_{i=1}^{n} sign(x - X_i) \right\| = 0 \qquad (2)$$

3. **Method**

We propose a novel clustering stopping rule based on two functions of $L_2$ Euclidean distance to the spatial median. The first one is the Average Between-Distances to the Medians (ABDM), where it is depending on calculating the average of $L_2$ distances between the medians of each pair of groups, and the second one is the Average Within -Distances to the Median (AWDM), which considers the average of $L_2$ distance of each individual observation to the spatial median of its group.

*3.1 Between and within distances to median ratio (BWDM)*

Suppose that $x_1, \ldots, x_n \in \Re^d$ is a $d$ – dimensional dataset with sample size n, for some cluster solution with total number of clusters K, let $k_i$ denotes the size of the i-th cluster, and $SM_i$ is spatial median of the i-th cluster considering the spatial median function defined (2), then the ABDM for multidimensional data can be defined as

$$\text{ABDM}(K) = \frac{1}{\binom{K}{2}} \sum_{\substack{i \neq j \\ j > i}}^{K} \| SM_i - SM_j \| = \frac{1}{\binom{K}{2}} \left( \sum_{\substack{i \neq j \\ j > i}}^{K} \sum (SM_i - SM_j)^2 \right)^{1/2} \qquad (3)$$

where $\binom{K}{2}$ is the total number of the possible combinations of each pair of clusters, $\|\cdot\|$ is the Euclidean norm. It is clear that the total number of possible combinations of the clusters' pairs is a monotonically increasing in the number of clusters K. Moreover, the difference between possible combinations in the solution (k=i) and solution (k=i-1) equals i-1. For instance, the total numbers of the possible combinations of each pair of clusters for k=6 is 15 pairs, while it is 10 when the number of clusters k=5, so the difference between them is 5 which is i-1 for i=6. Thus, ABDM serves as a measure of between-cluster variation, where it gives the biggest value for the most distinct clustering. Consequently, we can initially suggest that choosing the optimal number of clusters should be in accordance with the highest value of ABDM.

Similarly, AWDM serves as a measure of within-cluster variation, where it considers the distance between each observation in the cluster and its spatial median. Accordingly, we can define AWDM as:

$$\text{AWDM}(K) = \frac{1}{n}\sum_{i=1}^{K}\sum_{k=1}^{k_i}\|x_{ik} - SM_i\| \qquad (4)$$

where $\sum_{i=1}^{K} k_i = n$ is the total sample size. So, for each clustering solution (k), we get the corresponding averages ABDM and AWDM in order to calculate the ratio BWDM.

From (3) and (4), we proposed a Between and Within Distances to Median (BWDM) index as a ratio of ABDM to AWDM after assuming a specific denominator for each of them to account for number of observations and number of clusters. The denominators were specified following the same procedure of computing CH index that has been introduced by [22], where they considered the denominator of the between-variation as (k-1), and the denominator of the within-variation is the sample size minus the number of clusters (n-k). They showed that their criterion is analogous to the F-statistic in univariate analysis, where the degree of freedom is the same. Moreover, they pointed out that their criterion has been already used by [23] as an F-test in a multivariate cluster analysis. Accordingly, our new criterion BWDM index of clustering assignments over a given number of clusters K, is computed as following:

$$\text{BWDM}(K) = \frac{\text{ABDM}(K)/(K-1)}{\text{AWDM}(K)/(n-K)} \qquad (5)$$

In order to estimate the number of clusters $\widehat{K}$ based on the cluster-stopping rule BWDM, we should choose the value of K that maximizes the BWDM index value, such that:

$$\widehat{K} = \arg\max_{K \in \{2,\dots,K_{max}\}} \text{BWDM}(K) \qquad (6)$$

This method gives the highest value for the most distinct clustering, indicating strong homogeneity within the clusters and clear heterogeneity between clusters.

*3.2 The BWDM Algorithm*

Computation of the BWDM statistic proceeds as follows:

For each cluster solution:

(1) Calculate the average between-distances to the medians $\text{ABDM}(K)$ by applying equation (3).

(2) Calculate the average within-distances to the medians $\text{AWDM}(K)$ by applying equation (4).

(3) Calculate the ratio of the average between and within distances to the median $\text{BWDM}(K)$ by applying equation (5).

(4) Repeat steps (1) to (3) until K=Kmax.

(5) Choose the number of clusters K that maximizes the ratio $\text{BWDM}(K)$ as shown in equation (6).

## 4. Numerical Examples

In this section we evaluate the performance of our proposed clustering stopping rule algorithm using three simulated datasets and three well-known real-world datasets.

*4.1 Simulated data examples*

In these examples, we applied a series of simulated cases to assess the efficiency of the proposed algorithm. Three models have been simulated from a bivariate normal distribution while varying the mixture probabilities and considering different numbers of true clusters.

### 4.1.1 Simulated data 1

First, we generated data consisting of two groups from a bivariate mixture normal with mixture probabilities $(p_1, p_2)$ such that

$$x_1, \dots, x_n \sim p_1 N_2(\boldsymbol{\mu_1}, \boldsymbol{\Sigma}) + p_2 N_2(\boldsymbol{\mu_2}, \boldsymbol{\Sigma}) \qquad (7)$$

Where $p_1 = 0.7$, $p_2 = 0.3$, $N_2(\boldsymbol{\mu_1}, \boldsymbol{\Sigma})$ and $N_2(\boldsymbol{\mu_2}, \boldsymbol{\Sigma})$ denote a bivariate Gaussian with means $\boldsymbol{\mu_1} = \begin{pmatrix} 0 \\ 0 \end{pmatrix}$, $\boldsymbol{\mu_2} = \begin{pmatrix} 5 \\ 5 \end{pmatrix}$ and a covariance matrix $\boldsymbol{\Sigma} = I$. Second, we intentionally assumed that we have three groups to explore how the proposed method would respond under this scenario.

Figure (1) shows the scatter plot for a mixture of two groups from bivariate normal distribution considering the two cases K=2 and K=3. The BWDM($K = 2$) = 113.38, and BWDM($K = 3$) = 56.13, which suggests, depending on BWDM criterion, that $\widehat{K} = 2$ gives more distinct clustering than $\widehat{K} = 3$. In order to check the stability of BWDM over other potential numbers of clusters, we assumed that the maximum number of clusters allowed (Kmax) equals 10 (as frequently used in literature). As shown in Table 1 $\widehat{K} = 2$ maximizes the BWDM index. Figure 2 shows the BWDM curve for the assumed model, at Kmax=10, with the maximum point of BWDM(K) occurs at $\widehat{K} = 2$, which confirms the stability of our method.

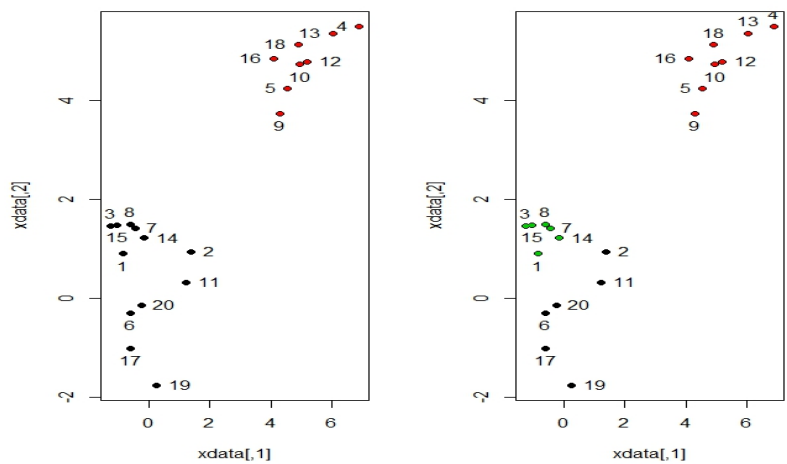

Figure1: Scatterplot for a mixture of three groups for simulated data 1 assuming 2 and 3 clusters

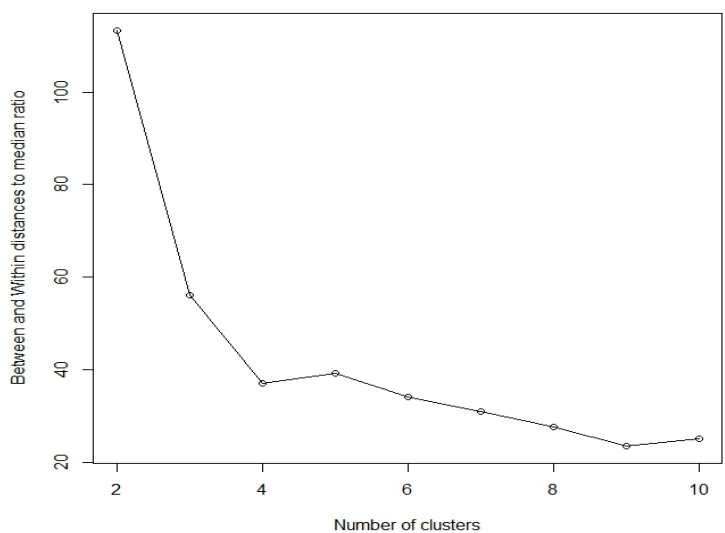

Figure 2: Between and within distances to median ratio (BWDM) curve for simulated data 1

### 4.1.2 Simulated data 2

In this scenario we generated data consists of three groups with mixture probabilities $(p_1, p_2, p_3)$ such that

$$x_1, \ldots, x_n \sim p_1 N_2(\boldsymbol{\mu_1}, \boldsymbol{\Sigma}) + p_2 N_2(\boldsymbol{\mu_2}, \boldsymbol{\Sigma}) + p_3 N_2(\boldsymbol{\mu_3}, \boldsymbol{\Sigma}) \tag{8}$$

Where $p_1 = p_2 = 0.3, p_3 = 0.4$, $N_2(\boldsymbol{\mu_1}, \boldsymbol{\Sigma})$, $N_2(\boldsymbol{\mu_2}, \boldsymbol{\Sigma})$ and $N_2(\boldsymbol{\mu_3}, \boldsymbol{\Sigma})$ denote a bivariate Gaussian with means $\boldsymbol{\mu_1} = \begin{pmatrix}0\\0\end{pmatrix}, \boldsymbol{\mu_2} = \begin{pmatrix}5\\5\end{pmatrix}, \boldsymbol{\mu_3} = \begin{pmatrix}10\\10\end{pmatrix}$ and a covariance matrix $\boldsymbol{\Sigma} = I$. In contrast to the first scenario, we intentionally supposed that we have two groups while the true number of clusters is 3. Figure 3 shows the scatterplot for a mixture of two groups from bivariate normal distribution considering the two cases K=2 and K=3. As reported in Table 1, the BWDM$(K = 2) = 61.03$, and BWDM$(K = 3) = 64.60$, suggesting that $\widehat{K} = 3$ gives more distinct clustering than $\widehat{K} = 2$ according to our proposed BWDM criterion. Moreover, we again examined the BWDM criterion for different numbers of clusters where Kmax=10. The BWDM curve (Figure 4) indicates that $\widehat{K} = 3$ is a reasonable choice as since K=3 represents indeed all the observations with more homogeneity within clusters and more heterogeneity between clusters compared to K=2 as shown in Figure 3.

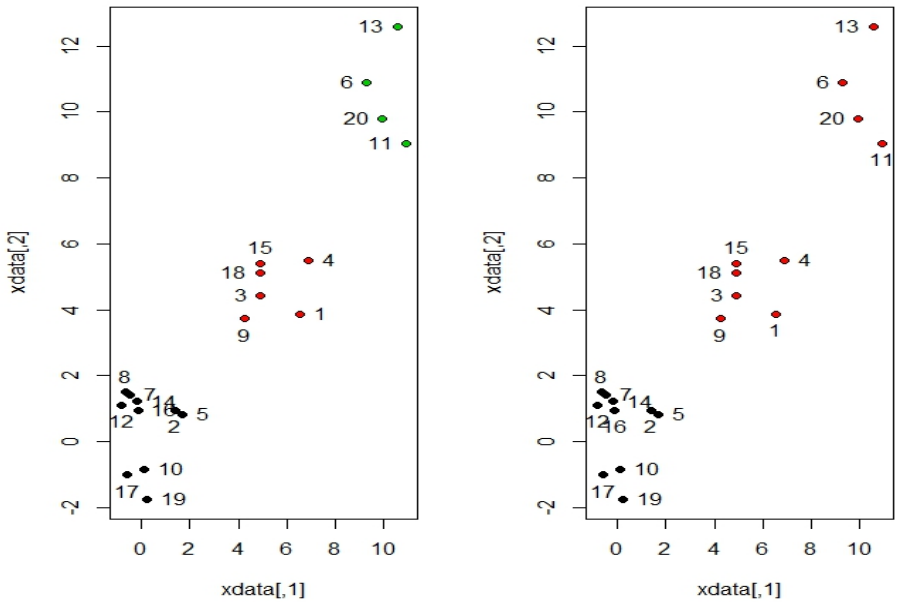

Figure 3: Scatterplot for a mixture of three groups for simulated data 2 assuming 2 and 3 clusters

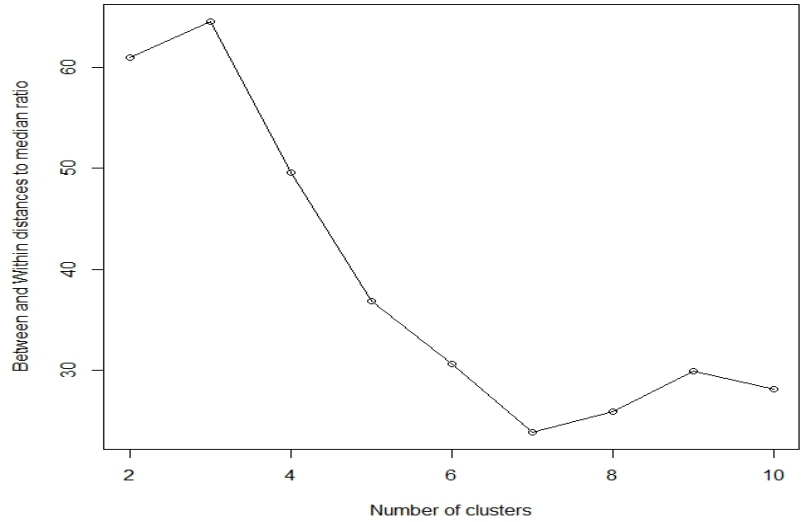

Figure 4: Between and within distances to median ratio (BWDM) curve for simulated data 2.

### 4.1.3 simulated data 3

In this data we consider a model consists of a mixture of four bivariate normal distributions where the weights are assumed to be equal ($p_i = 0.25; i = 1, \dots, 4$) such that:

$$x_1, \dots, x_n \sim p_1 N_2(\boldsymbol{\mu_1}, \boldsymbol{\Sigma}) + p_2 N_2(\boldsymbol{\mu_2}, \boldsymbol{\Sigma}) + p_3 N_2(\boldsymbol{\mu_3}, \boldsymbol{\Sigma}) + p_4 N_2(\boldsymbol{\mu_4}, \boldsymbol{\Sigma}) \qquad (9)$$

Where $N_2(\boldsymbol{\mu_1}, \boldsymbol{\Sigma})$, $N_2(\boldsymbol{\mu_2}, \boldsymbol{\Sigma})$, $N_2(\boldsymbol{\mu_3}, \boldsymbol{\Sigma})$ and $N_2(\boldsymbol{\mu_4}, \boldsymbol{\Sigma})$ denote a bivariate Gaussian with means $\boldsymbol{\mu_1} = \begin{pmatrix} 0 \\ 0 \end{pmatrix}$, $\boldsymbol{\mu_2} = (4)$, $\boldsymbol{\mu_3} = \begin{pmatrix} -4 \\ 4 \end{pmatrix}$ and a covariance matrix $\boldsymbol{\Sigma} = I$

Similar to the previous scenarios, we intentionally assumed that we have a varied number of clusters different from the true number of four clusters with maximum number of clusters =10. Figure 5 shows the scatter plot for a mixture of four groups from bivariate normal distribution with equal probabilities. As shown in Table 1 and Figure 6, $\widehat{K} = 4$ maximizes the BWDM index which confirms the stability of our method.

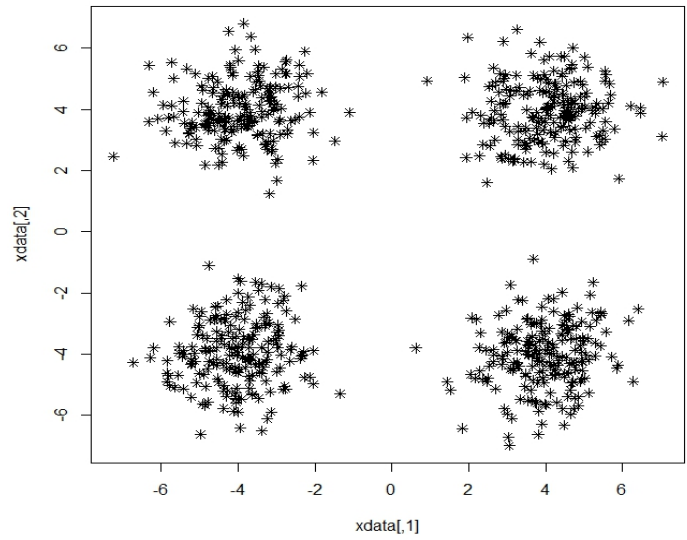

Figure 5: Scatterplot for a mixture of four bivariate normal distributions with equal weights and sample sizes.

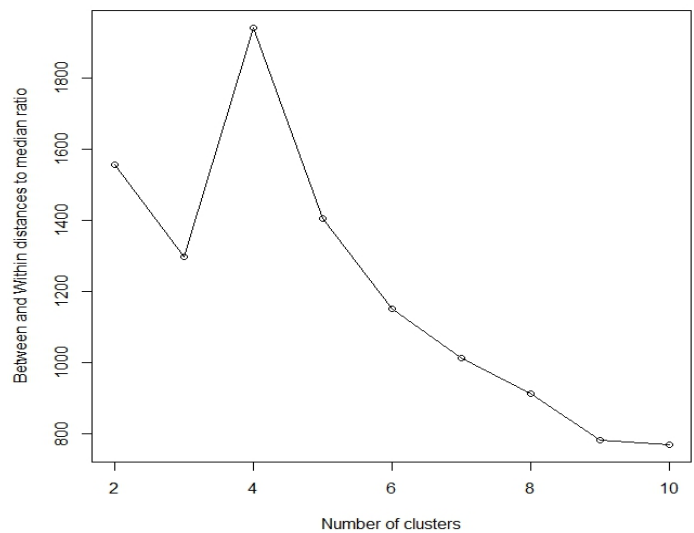

Figure 6: Between and within distances to median ratio (BWDM) curve for simulated data 3.

## 4.2 Real-world datasets examples

In order to understand how well our algorithm works in practical situations, we evaluated the proposed clustering stopping rule algorithm using three well-known real datasets in statistics and data science including; Old Faithful data, Financial data, and Iris data.

### 4.2.1 Real data 1: Old Faithful Data

This dataset includes two variables, the waiting time between eruptions, and the duration of the eruption in minutes for the old faithful geyser in Yellowstone National Park, Wyoming, USA [24]. As shown in the scatter plot of the faithful dataset (Figure 7), this dataset clearly consists of two clusters, the short and the long eruptions. Consistently, results of our proposed stopping rule algorithm presented in Figure 8 and Table 1 showed that $\widehat{K} = 2$ maximizes the BWDM index, suggesting that our proposed method performs well in this dataset.

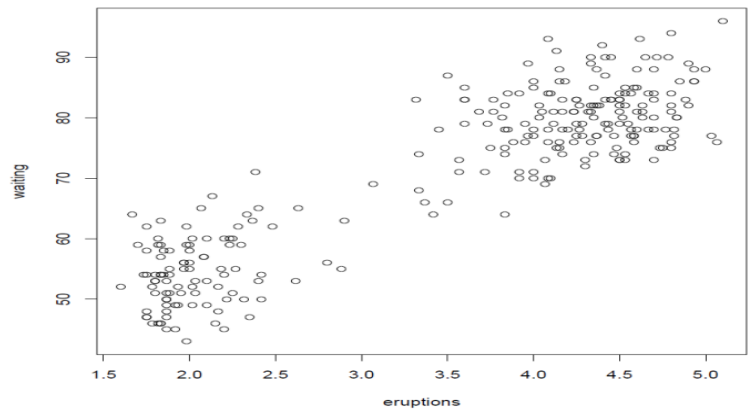

Figure 7: Scatter plot for Old Faithful Data

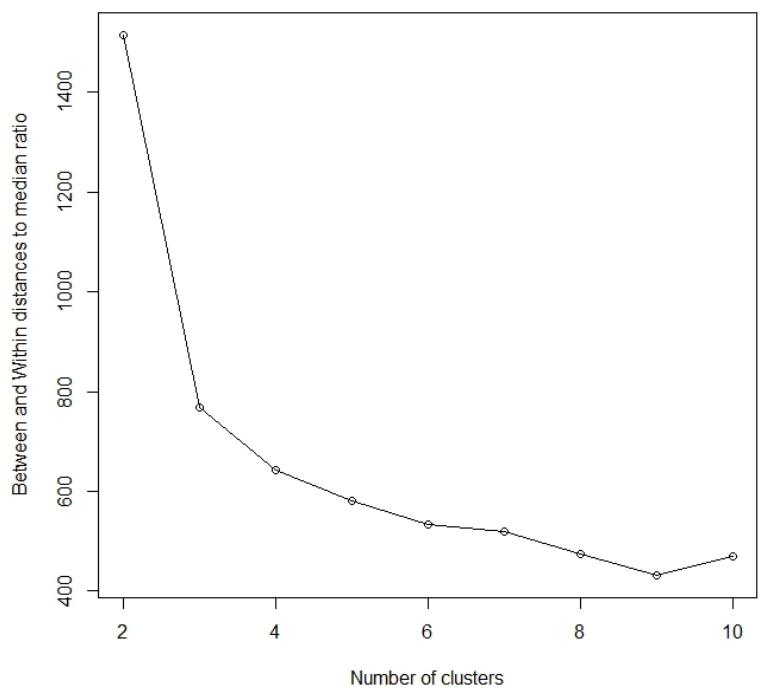

Figure 8: Between and within distances to median ratio (BWDM) curve for Old Faithful dataset.

### 4.2.2 Real data 2: Financial Data

This dataset includes measurements of three variables that monitor the short term and medium term performance and medium term volatility of investment funds operating in Italy since April 1996 (Table A.16 of [25]). The financial data consists of two clusters representing two different kinds of funds; stock funds and balanced funds [26] presented by a 3D scatter plot (Figure 9). Our proposed method performs well in this dataset as well where $\hat{K} = 2$ maximizes the BWDM index as shown in Figure 10 and Table 1.

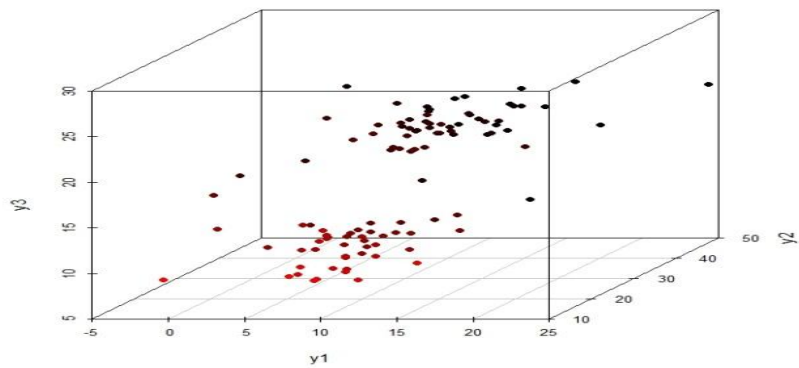

Figure 9: 3D scatterplot for Financial data

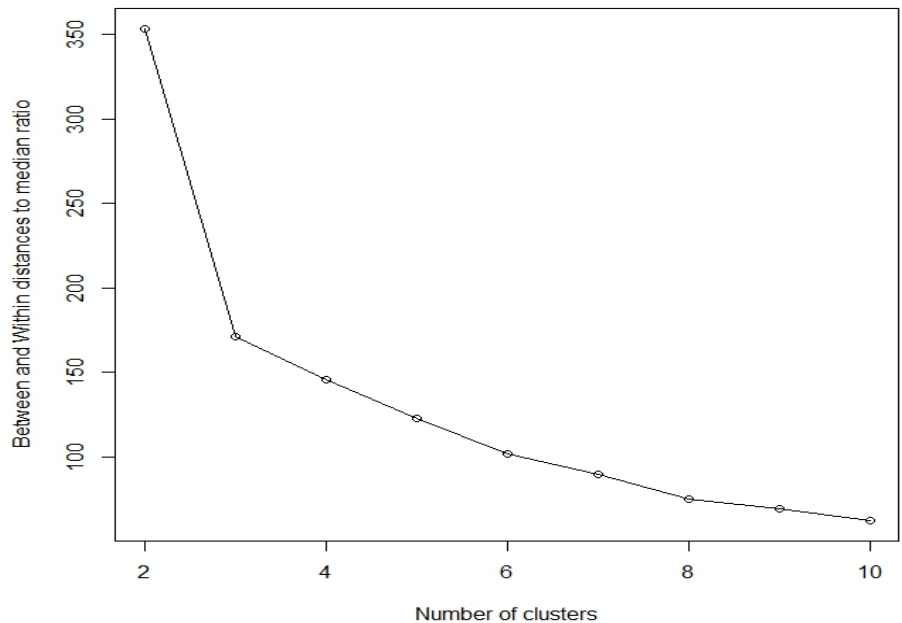

Figure 10: Between and within distances to median ratio (BWDM) curve for Financial data

### 4.2.3 Real data 3: Iris Data

This dataset consists of four variables representing length and width of Sepal and Petal. Although this data contains three different types of irises (Setosa, Versicolour, and Virginica), most clustering techniques consider them as two groups since iris Virginica and iris Versicolour are not separable without the species information that Fisher used, as shown in Figure 11. Table 1 and Figure 12 showed that, our proposed algorithm continued to successfully determine the true number of clusters where $\widehat{K} = 2$ maximizes the BWDM index.

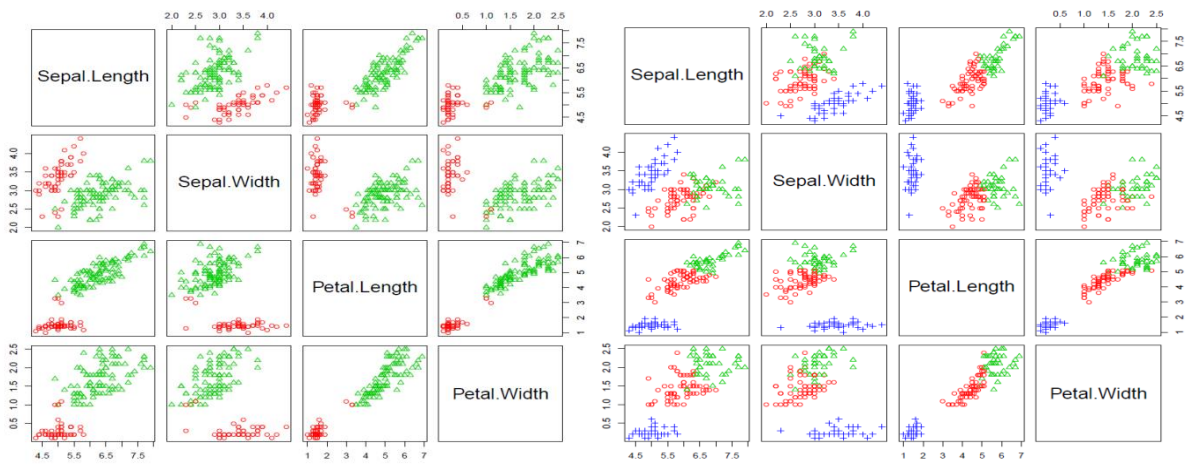

Figure 11: Scatter plot for Iris data assuming two and three clusters

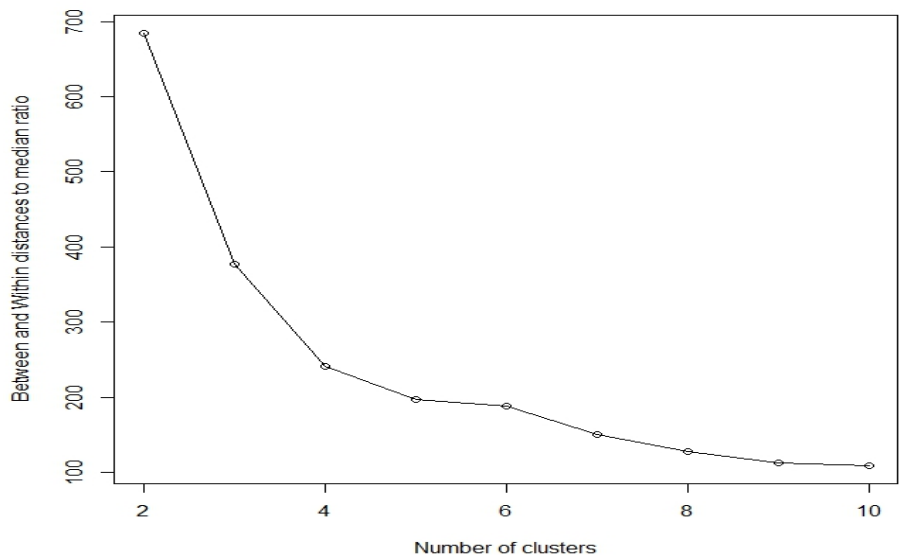

Figure 12: Between and within distances to median ratio (BWDM) curve for Iris data

## 5. Comparison with other clustering algorithms

In this section we considered the three real data sets mentioned above to compare the performance of our method with other 13 clustering algorithms in order to assess its efficacy in determining the number of true clusters. The determined number of clusters for the algorithms used for comparison are presented in Table 2 along with our proposed algorithm (BWDM). These algorithms were applied to the Old Faithful data, the Financial data, and the Iris data. As shown in Table 2, five of the 13 algorithms (GMM (mclust), K-means, DBSCAN, KMD, and densityClust) did not identify the correct number of clusters for Old Faithful data. For Financial data, 6 algorithms (GMM (mclust), HDDC, DBSCAN, DDC, SNN, and densityClust) suggested wrong number of clusters while for the Iris data, only 5 algorithms were successful in identifying the correct number of clusters (GMM, PAM, DBSCAN, WSR, and densityClust). However, in addition to our algorithm, only two other algorithms were stable in identifying the correct number of clusters for all considered datasets (PAM and WSR) while most algorithms (GMM, K-means, HDDC, DBSCAN, KMD, DDC, SNN, and densityClust) identified the correct number of clusters for only one of the three datasets.

## 6. Concluding remarks

This study presents a clustering stopping-rule algorithm for determining the optimal number of clusters, based on the spatial median. We proposed an algorithm that aims to balance the trade-off between intra-cluster homogeneity and inter-cluster heterogeneity. The main idea is maximizing the ratio of between-cluster to within-cluster variation, while accounting for both the number of clusters and observations. As a nonparametric approach, it is robust against distributional assumptions. In addition, as it is a spatial median-based approach it performs well under the presence of outliers. This algorithm has been validated through simulations where we considered three different scenarios of data generated from bivariate normal distribution. We further assessed its stability and effectiveness using three real-world datasets: the Old Faithful data, Financial data, and Iris data. The proposed algorithm successfully identified the correct number of clusters for all datasets. In addition, we compared its performance to 13 traditional clustering algorithms, demonstrating that our approach outperformed 11 of these in determining the number of clusters which highlights the method's reliability for clustering in multivariate data. several directions can be pursued to extend the algorithm's applicability. One potential area is the exploration of more complex clustering structures, such as overlapping clusters. A further possible extension is adapting the algorithm to handle more complex datasets such as functional datasets. This extension could lead to significant contributions in areas such as time series analysis and further practical values in various fields such as medical imaging, or environmental studies, where functional data types are prevalent.

Table 1: Between and within distances to median ratio (BWDM)

|  | Simulated Data | | | | | | | | |
|---|---|---|---|---|---|---|---|---|---|
| K | 2 | 3 | 4 | 5 | 6 | 7 | 8 | 9 | 10 |
| **Simulated data1** | 113.38 | 56.13 | 37.10 | 39.33 | 34.20 | 30.99 | 27.58 | 23.48 | 25.10 |
| **Simulated data2** | 61.03 | 64.60 | 49.58 | 36.81 | 30.60 | 23.82 | 25.93 | 29.86 | 28.14 |
| **Simulated data3** | 1554.70 | 1296.54 | 1940.96 | 1403.62 | 1149.28 | 1010.36 | 910.49 | 781.04 | 767.38 |
|  | **Real-World Datasets** | | | | | | | | |
| **Real data 1** | 1516.09 | 767.43 | 642.35 | 581.33 | 533.80 | 519.28 | 473.76 | 430.85 | 470.21 |
| **Real data 2** | 353.53 | 171.58 | 145.52 | 122.62 | 101.98 | 89.56 | 75.44 | 69.41 | 62.30 |
| **Real data 3** | 684.95 | 377.18 | 241.28 | 196.46 | 188.34 | 150.59 | 127.31 | 112.85 | 108.59 |

Table 2: Comparison of different algorithm used to determine number of clusters for Old Faithful, Financial, and Iris datasets.

| Method | Criteria used to determine number of clusters | Data | | |
|---|---|---|---|---|
| | | Old Faithful data | Financial data | Iris data |
| **BWDM** | Considers the variation of between and within clusters. | 2 | 2 | 2 |
| **GMM (mclust) [27]** | Depends on Bayesian information criterion (BIC). | 3 | 3 | 2 |
| **K-means [28]** | Uses the Calinski–Harabasz (CH) index. | 10 | 2 | 3 |
| **HDDC [29]** | Uses Bayesian information criterion (BIC). | 2 | 3 | 3 |
| **MixtPPCA [30]** | Uses Bayesian information criterion (BIC). | 2 | 2 | 3 |
| **PAM [31]** | Depends on the optimum average silhouette width. | 2 | 2 | 2 |
| **DBSCAN [32]** | Uses a density-based approach to identify regions of high density. | 3 | 1 | 2 |
| **WSR [11]** | A Weighted Spatial Ranks based approach. | 2 | 2 | 2 |
| **KMD [33]** | Depends on K-medoids algorithm. | 3 | 2 | 3 |
| **FCM [34]** | A fuzzy C-means clustering algorithm. | 2 | 2 | 3 |
| **GG[35]** | The Gath–Geva clustering algorithm. | 2 | 2 | 3 |
| **DDC [36]** | Distance density clustering. | 2 | 3 | 3 |
| **SNN [37]** | Clustering with shared nearest neighbour clustering. | 2 | 3 | 3 |
| **densityClust [38]** | Clustering by fast search and finding of density peaks. | 1 | 1 | 2 |

**Data Availability**

The data used to support the findings of the study are available in the public domain and are appropriately referenced in this article.

**Conflicts of Interest**

The authors declare that they have no conflicts of interest.

**Funding**

This research did not receive any specific grant from funding agencies in the public, commercial, or not-for-profit sectors.

References

1. Wedel, M. and W.A. Kamakura, *Market segmentation: Conceptual and methodological foundations*. 2000: Springer Science & Business Media.
2. Milligan, G.W. and M.C. Cooper, *An examination of procedures for determining the number of clusters in a data set.* Psychometrika, 1985. **50**: p. 159-179.
3. Rule, A.S., *Determination of the Number of Clusters in a Data Set.* 2013.
4. Baragilly, M., H. Gabr, and B.H. Willis, *Clustering functional data using forward search based on functional spatial ranks with medical applications.* Statistical Methods in Medical Research, 2022. **31**(1): p. 47-61.
5. Li, M.J., et al., *Agglomerative fuzzy k-means clustering algorithm with selection of number of clusters.* IEEE transactions on knowledge and data engineering, 2008. **20**(11): p. 1519-1534.
6. Liao, H.-y. and M.K. Ng, *Categorical data clustering with automatic selection of cluster number.* Fuzzy Information and Engineering, 2009. **1**(1): p. 5-25.
7. Kaufman, L. and P.J. Rousseeuw, *Finding groups in data: an introduction to cluster analysis*. 2009: John Wiley & Sons.
8. Roberts, S.J., *Parametric and non-parametric unsupervised cluster analysis.* Pattern Recognition, 1997. **30**(2): p. 261-272.
9. Godichon-Baggioni, A. and S. Surendran, *A Penalized Criterion for Selecting the Number of Clusters for K-Medians.* Journal of Computational and Graphical Statistics: p. 1-12.
10. Möttönen, J. and H. Oja, *Multivariate spatial sign and rank methods.* Journaltitle of Nonparametric Statistics, 1995. **5**(2): p. 201-213.
11. Baragilly, M.H., H. Gabr, and B.H. Willis, *Clustering Analysis of Multivariate Data: A Weighted Spatial Ranks-Based Approach.* Journal of Probability and Statistics, 2023. **2023**(1): p. 8849404.
12. Sirkiä, S., et al., *Tests and estimates of shape based on spatial signs and ranks.* Journal of Nonparametric Statistics, 2009. **21**(2): p. 155-176.
13. Brown, B.M., *Statistical uses of the spatial median.* Journal of the Royal Statistical Society Series B: Statistical Methodology, 1983. **45**(1): p. 25-30.
14. Kemperman, J., *The median of a finite measure on a Banach space.* Statistical data analysis based on the L1-norm and related methods (Neuchâtel, 1987), 1987: p. 217-230.

15. Chaudhuri, P., *Multivariate location estimation using extension of R-estimates through U-statistics type approach.* The Annals of Statistics, 1992: p. 897-916.
16. Chakraborty, B., P. Chaudhuri, and H. Oja, *Operating transformation retransformation on spatial median and angle test.* Statistica Sinica, 1998: p. 767-784.
17. Chakraborty, B. and P. Chaudhuri, *On a transformation and re-transformation technique for constructing an affine equivariant multivariate median.* Proceedings of the American mathematical society, 1996. **124**(8): p. 2539-2547.
18. Hettmansperger, T.P. and R.H. Randles, *A practical affine equivariant multivariate median.* Biometrika, 2002. **89**(4): p. 851-860.
19. Vardi, Y. and C.-H. Zhang, *A modified Weiszfeld algorithm for the Fermat-Weber location problem.* Mathematical Programming, 2001. **90**: p. 559-566.
20. Baragilly, M.H.H., *Clustering multivariate and functional data using spatial rank functions*. 2016, University of Birmingham.
21. Zuo, Y., *Multidimensional medians and uniqueness.* Computational Statistics & Data Analysis, 2013. **66**: p. 82-88.
22. Caliński, T. and J. Harabasz, *A dendrite method for cluster analysis.* Communications in Statistics-theory and Methods, 1974. **3**(1): p. 1-27.
23. Edwards, A.W. and L.L. Cavalli-Sforza, *A method for cluster analysis.* Biometrics, 1965: p. 362-375.
24. Azzalini, A. and A.W. Bowman, *A look at some data on the Old Faithful geyser.* Journal of the Royal Statistical Society: Series C (Applied Statistics), 1990. **39**(3): p. 357-365.
25. Atkinson, A.C., M. Riani, and A. Cerioli, *Exploring multivariate data with the forward search*. 2013: Springer Science & Business Media.
26. Baragilly, M. and B. Chakraborty. *Determining the number of clusters using multivariate ranks*. in *Recent Advances in Robust Statistics: Theory and Applications*. 2016. Springer.
27. Fraley, C. and A.E. Raftery, *Model-based clustering, discriminant analysis, and density estimation.* Journal of the American statistical Association, 2002. **97**(458): p. 611-631.
28. Harabasz, C.T. and M. Karoński, *A dendrite method for cluster analysis*, in *Communications in Statistics*. 1974. p. 1-27.
29. Bouveyron, C., S. Girard, and C. Schmid, *High-dimensional data clustering.* Computational statistics & data analysis, 2007. **52**(1): p. 502-519.
30. Tipping, M.E. and C.M. Bishop, *Mixtures of probabilistic principal component analyzers.* Neural computation, 1999. **11**(2): p. 443-482.
31. Reynolds, A.P., et al., *Clustering rules: a comparison of partitioning and hierarchical clustering algorithms.* Journal of Mathematical Modelling and Algorithms, 2006. **5**: p. 475-504.
32. Ester, M., et al. *A density-based algorithm for discovering clusters in large spatial databases with noise*. in *kdd*. 1996.
33. Park, H.-S. and C.-H. Jun, *A simple and fast algorithm for K-medoids clustering.* Expert systems with applications, 2009. **36**(2): p. 3336-3341.
34. Torabi, A.J., et al., *Application of clustering methods for online tool condition monitoring and fault diagnosis in high-speed milling processes.* IEEE Systems Journal, 2015. **10**(2): p. 721-732.
35. Yu, K., T.R. Lin, and J.W. Tan, *A bearing fault diagnosis technique based on singular values of EEMD spatial condition matrix and Gath-Geva clustering.* Applied Acoustics, 2017. **121**: p. 33-45.

36. Ma, R. and R. Angryk. *Distance and density clustering for time series data*. in *2017 IEEE international conference on data mining workshops (ICDMW)*. 2017. IEEE.
37. Ertöz, L., M. Steinbach, and V. Kumar. *Finding clusters of different sizes, shapes, and densities in noisy, high dimensional data*. in *Proceedings of the 2003 SIAM international conference on data mining*. 2003. SIAM.
38. Rodriguez, A. and A. Laio, *Clustering by fast search and find of density peaks.* science, 2014. **344**(6191): p. 1492-1496.